# DRAWING FROM HATS BY NOISE-BASED LOGIC

BRUCE ZHANG[1], LASZLO BELA KISH[1], CLAES-GÖRAN GRANQVIST[2]

*(1) Department of Electrical Engineering, Texas A&M University, College Station, TX 77843-3128, USA; laszlo.kish@ece.tamu.edu*

*(2) Department of Engineering Sciences, The Ångström Laboratory, Uppsala University, P.O. Box 534, SE-75121 Uppsala, Sweden*

# DRAWING FROM HATS BY NOISE-BASED LOGIC

We utilize the asymmetric random telegraph wave-based instantaneous noise-base logic scheme to represent the problem of drawing numbers from a hat, and we consider two identical hats with the first $2^N$ integer numbers. In the first problem, Alice secretly draws an arbitrary number from one of the hats, and Bob must find out which hat is missing a number. In the second problem, Alice removes a known number from one of the hats and another known number from the other hat, and Bob must identify these hats. We show that, when the preparation of the hats with the numbers is accounted for, the noise-based logic scheme always provides an exponential speed-up and/or it requires exponentially smaller computational complexity than deterministic alternatives. Both the stochasticity and the ability to superpose numbers are essential components of the exponential improvement.

## 1. Introduction

Noise-based logic (NBL) [1-8] was inspired by the human brain, which uses stochastic signals to carry logic information. We first introduce the NBL concept and the specific scheme to be utilized for the solution of our problem solution.

### *1.1. On noise-based logic*

Independent stochastic processes (noises) with zero mean are uncorrelated, that is, orthogonal [1], and this is true also for their products. In NBL, the logic information is carried by independent (orthogonal) stochastic processes (reference noises) of zero mean which form the basis that represents an *N*-dimensional vector space [1] and a $2^N$ dimensional product-string-based hyperspace [2]. The NBL gates are driven by the input logic information and the reference noises; see Figure 1. The logic information is carried by these noises and their superposition, by the products of these noises, and by the superposition of these products forming the logic *hyperspace* of $2^N$ dimension [1,2].

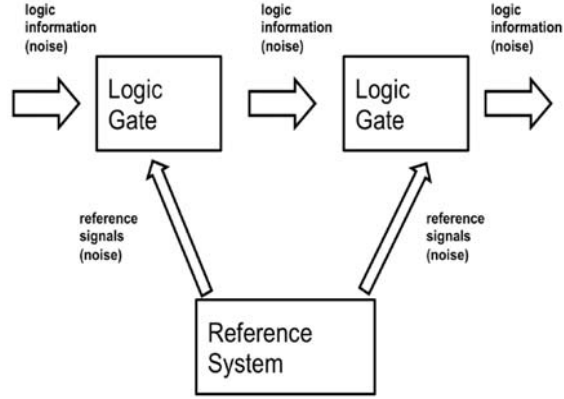

Fig. 1. Generic noise-based logic hardware scheme. In instantaneous noise-based logic, the logic gates do not use statistical correlators or time averaging [4]. Instead algebraic operations, or coincidence detection, are used to instantaneously create the output state (signal).

Instantaneous noise-based logic (INBL) is a recently proposed [3-7] class of NBL wherein no statistical correlators and/or time averaging are used. The hyperspace has a good potential for performing large-scale special-purpose operations with exponential speed-up and low hardware complexity.

One class of stochastic time functions that can effectively be used in INBL is periodic-time random telegraph waves (RTWs). The amplitude of the simplest (symmetric) RTW randomly changes its sign at the beginning of each clock period.

*1.2. The asymmetric random telegraph wave based INBL [5]*

To represent a binary bit in this system, we use two independent RTWs, $H_i(t)$ and $L_i(t)$, for the High- and Low-bit values, respectively, where *i* is bit-index and *t* is time. Such a pair of RTWs is called a *noise-bit* [1-7]. In this paper, we use the *asymmetric scheme* [5] characterized by $|H_i(t)| \neq |L_i(t)|$ with $|H_i(t)| = 1$ and $|L_i(t)| = 0.5$; see Figures 2 and 3.

A bit-string is the product of the RTWs, denoted $G_j^{(k)}(t)$, corresponding to the bit-values in the string. Such a bit-string, $\prod_{j=1}^{N} G_j^{(k)}(t)$, represents a binary integer number of *N*-bit resolution, where *j* is the index of the *j*-th noise bit and *k* is the number that is represented. In an *N* noise-bit system, there are $2^N$ different product strings representing $2^N$ different integer numbers so that $0 \leq k < 2^{N-1}$. The noise in Figure 4 is an example signifying the number 2 (binary number 00000000000000000000000000000010) in a 32 noise-bits INBL scheme. In an ordinary computer one cannot superimpose bits and numbers in a single wire, but the INBL scheme allows for "physical" superposition of these numbers.

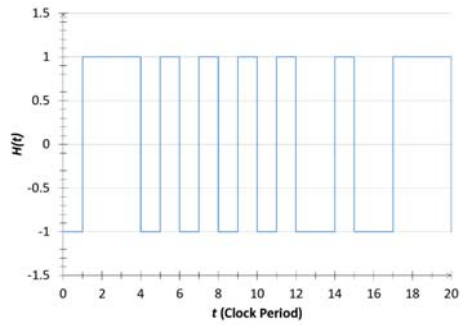

Fig. 2. Computer simulation of a random telegraph wave carrying the High-bit value of a noise-bit in the asymmetric INBL scheme.

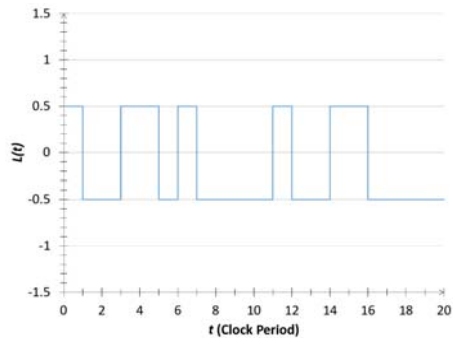

Fig. 3. Computer simulation of a random telegraph wave carrying the Low-bit value of a noise-bit in the asymmetric INBL scheme.

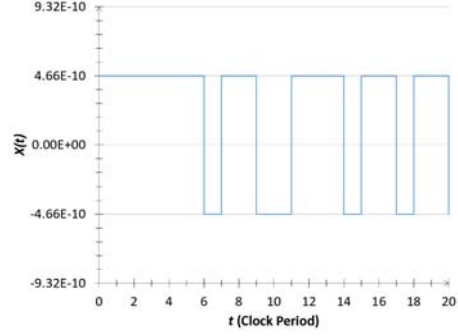

Fig. 4. Computer simulation of the product-string $\prod_{j=1}^{N} G_j^{(2)}(t)$ of the number 2 (binary number 00000000000000000000000000000010) in a $N = 32$ noise-bits INBL scheme.

The superposition of all $2^N$ numbers is called a "universe" [2]. Even though the universe $U(t)$ has an exponentially large number ($2^N$) of elements, it can be produced in INBL by $2N - 1$ operations by use of the Achilles ankle method [2], the result being

$$U(t) = \prod_{i=1}^{N} [H_i(t) + L_i(t)]. \qquad (1)$$

Figure 5 shows an example, a computer-generated 32-bits universe (see Equation 1), which is the superposition of over four billion binary numbers of 32-bit resolution.

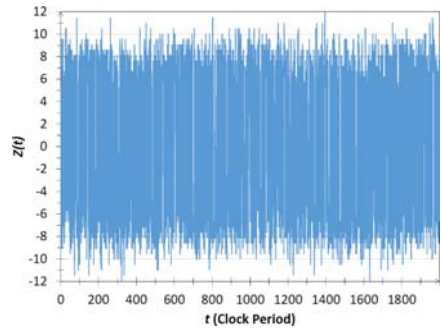

Fig. 5. Computer simulation of the $N = 32$ noise-bit universe in the asymmetric INBL scheme over 2000 clock periods. The time function $U(t)$ of the universe, which represents the superposition of all integer numbers from 0 to 4294967295, is plotted in a logarithmically distorted way, defined by $Z(t) = \text{sign}[U(t)] \log[2^{32}|U(t)|]$, for better visibility of small and large variations.

In this paper, we apply the above INBL scheme to the problem of drawing numbers from a hat and to identify such hats.

## 2. Drawing one arbitrary number from one of two hats

### 2.1. Set-up

Our goal is to construct a model that reflects the physical nature of the problem in the sense that:

i) There must be a measure, which can be deterministic or statistical, that yields a strictly monotonic relation between the number of numbers in the hat and that particular measure of their set.

ii) The numbers can be identified, separated or taken out of the set.

After these remarks, our system is as follows. To represent all possible $N$-bit integer numbers in the interval $[0, 2^N - 1]$, Alice creates $N$ noise-bits, that is, $2N$ orthogonal RTWs in the asymmetric fashion described above. Next, Alice creates two independent and identical hats containing the same universe by

$$U_1(t) = U_2(t) = \prod_{i=1}^{N} \left[ H_i(t) + L_i(t) \right] \;, \tag{2}$$

where indices 1 and 2 denote Hat-1 and Hat-2, respectively. Figure 5 refers to a realization with 32 noise-bits.

### 2.2. Drawing

Alice randomly draws an arbitrary number from one of the two hats by subtracting the time function of the drawn number's product-string from the relevant universe. As an example, suppose that the number $k$ was drawn from Hat-1. The superposition (universe) $U_1(t)$ of the first hat is then reduced to the superposition

$$U_1^{(k)}(t) = \prod_{i=1}^{N}[H_i(t)+L_i(t)] - \prod_{j=1}^{N} G_j^{(k)}(t) , \qquad (3)$$

where the $G_j^{(k)}(t)$ represents the RTW of the value of the $j$-th bit of the number $k$ $\left(G_j^{(k)}(t) \in \{H_j(t), L_j(t)\}\right)$. Figure 6 illustrates the case of $N = 32$ and $k = 2$. At the same time, the superposition in Hat-2 maintains its original universe $U_2(t)$.

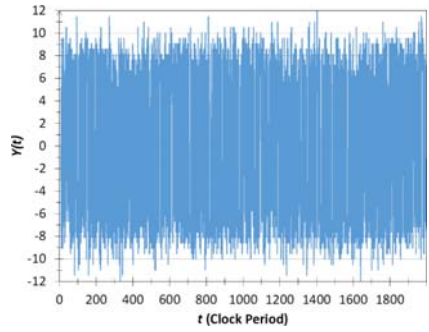

Fig. 6. Computer simulation of the $N = 32$ noise-bit universe shown in Figure 5 reduced by the product-string of the missing number $p = 2$ (see Figure 4), where the time function $U^{(p)}(t)$ is plotted in a logarithmically distorted way, defined by $Y(t) = \text{sign}[U^{(p)}(t)] \log[2^{32}|U^{(p)}(t)|]$, for better visibility of small and large variations.

## 2.3. Decision scheme

Bob's goal is now to identify which hat is missing the number $k$. Bob generates the universe in the original hats (see Equation 2) and then compares it to the actual superpositions in each of the hats. The matching superposition will identify the hat with the intact superposition, that is, with all numbers. Mathematically speaking (i) Bob generates the universe

$$U_{\text{Bob}}(t) = \prod_{i=1}^{N}\left[H_i(t) + L_i(t)\right], \qquad (4)$$

and (ii) then he can compare the time function of his superposition with those of the hats. From Equations 2 and 3, the differences of $U_{\text{Bob}}(t)$ and $U_1(t)$, and of $U_{\text{Bob}}(t)$ and $U_2(t)$, are

$$U_{\text{Bob}}(t) - U_1(t) = 0, \qquad (5)$$

and

$$U_{\text{Bob}}(t) - U_2(t) = \prod_{j=1}^{N} G_j^{(k)}(t) \neq 0. \qquad (6)$$

However, it is enough to carry out one of these comparisons. If the difference of Bob's superposition and that of a chosen hat is zero, then the number $k$ is missing from that hat; see Equation 5. If the difference is non-zero (see Equation 6) then the hat has all of the numbers, and the number is missing from the other hat.

Finally we note that, using the Stacho method [8], even the value $k$ of the number can be identified via $O_t(N \ln N)$ time steps from the RTW signal $\prod_{j=1}^{N} G_j^{(k)}(t)$ obtained by Bob's operation (see Equation 5); however this is not the subject of our present work.

*2.4. Computational complexity*

If the scheme outlined above is realized via special-purpose hardware, the hardware complexity of the set-up, drawing, and decision schemes is $O_h(N^2)$ because the parallel hardware realization of Equations 2–6 requires a hardware size scaling with $N^2$

, where one factor *N* comes from the bit resolutions and the another factor *N* emanates from the number of calculation steps. Such a protocol has $O_t(1)$ time complexity, because the decision about the hats is instantaneous (see Equations 5 and 6); it takes place during the first clock steps.

In conclusion, the *computation complexity* of the noise-based logic realization of the number drawing with hat identification is $O(N^2)$.

## 3. Drawing two numbers from two hats

### 3.1. Set-up

Alice sets up the two independent and identical hats as before; see Equation 2. Then she selects two random numbers, *p* and *q*, from the set $\{1,...,2^N\}$ and informs Bob about them.

### 3.2. Drawing

Alice secretly removes one of these numbers from Hat-1 and the other number from Hat-2, in the same way as earlier done; see Equation 3. Suppose she removes the number *p* from Hat-1 and *q* from Hat-2; then the remaining superpositions are

$$U_1^{(p)}(t) = \prod_{i=1}^{N}[H_i(t)+L_i(t)] - \prod_{j=1}^{N} G_j^{(p)}(t) \, , \tag{7}$$

and

$$U_2^{(q)}(t) = \prod_{i=1}^{N}[H_i(t)+L_i(t)] - \prod_{j=1}^{N} G_j^{(q)}(t) \, . \tag{8}$$

### 3.3. Decision

Bob proceeds in the same manner as before. He also generates the superpositions

$$U^{(p)}(t) = \prod_{i=1}^{N}[H_i(t)+L_i(t)] - \prod_{j=1}^{N} G_j^{(p)}(t) \tag{9}$$

and

$$U^{(q)}(t) = \prod_{i=1}^{N}\left[H_i(t) + L_i(t)\right] - \prod_{j=1}^{N} G_j^{(q)}(t) \tag{10}$$

and then compares the superpositions in Equations 7 and 8 with the superpositions in Equations 9 and 10. As above, such a comparison is made easy by producing the differences of the superpositions in the hats and the superpositions created by Bob. When there is a match, the result is uniformly zero. Otherwise the result is the difference of the time functions of the product-strings representing the two numbers. For example, in the above case, if Bob tests the first hat (Equation 7) with his second superposition (Equation 10) he gets

$$U_1^{(p)}(t) - U^{(q)}(t) = \prod_{j=1}^{N} G_j^{(q)}(t) - \prod_{j=1}^{N} G_j^{(p)}(t) , \tag{11}$$

which cannot always be zero as shown below.

### 3.4. Computational complexity

If the number of zero-bit values in the binary $N$-bit representation of numbers $p$ and $q$ are $r$ and $s$, respectively, the asymmetric RTW system (see Section 1.2) yields $\left|\prod_{j=1}^{N} G_j^{(p)}(t)\right| = 2^{-r}$ and $\prod_{j=1}^{N} G_j^{(q)}(t) = 2^{-s}$. Thus, in the cases when $r \neq s$, the difference of non-matching superpositions (e.g. Equation 11) will never be zero. In these situations, the decision is instantaneous so that its time complexity is $O_t(1)$.

However, in the special case of $r = s$, the difference of non-matching superpositions (e.g. Equation 11) is zero with probability $P(1) = 0.5$. The probability that the difference of two non-matching superpositions is zero over $n$ clock steps is $P(n) = 2^{-n}$; see Figure 7 and Reference [9]. If Bob is simultaneously testing the differences of matching and non-matching superpositions when both differences give zero values, then the situation is not conclusive. If there is a difference between the superpositions, the error in the decision is zero. However, the special case of *r* = *s*, with the exponentially decaying probability in time of not being able to make the decision, does not weaken the $O_t(1)$ time complexity of the decision because *p(n)* is independent of *N*. Moreover, the exponential cut-off of *p(n)* results in $(1-0.5)^{-1} = 2$ expected value of clock steps (the expected value of decision time due to the corresponding geometrical series), which also implies $O_t(1)$ time complexity. As in the first problem, see Section 2.4, the hardware complexity of the set-up, drawing, and decision schemes is $O_h(N^2)$ because the number of elements/operations scales linearly with *N* and each one handles *N* bit numbers.

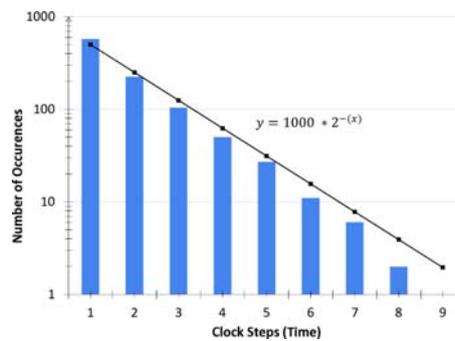

Fig. 7. Computer simulated distribution of number of clock periods required to observe the difference of signals for error-free response. The same *N* = 32 noise bit system was used as in Figures 4–6. The numbers drawn were *p* = 2 and *q* = 1, that is, binary numbers 00000000000000000000000000000010 and 00000000000000000000000000000001, respectively.

In conclusion, the *computation complexity* of the noise-based logic realization of the double-number drawing with hat identification problem is $O(N^2)$.

## 4. Conclusions

The two drawing and decision schemes discussed in this paper require polynomial computational complexity when realized with the hyperspace of instantaneous noise-based logic. Thus, with these schemes, noise-based logic offers an *exponential speed-up of computation* (or exponential reduction of computational complexity) compared to the case of classical computational schemes when the complexity requirements of the set-up are accounted for. One of the fastest classical set-up can involve a real hat with numbers of identical or different weights and the decision, which is instantaneous, can be done by a weighing scale.

**Acknowledgements**

Discussions with Tamas Horvath, Laszlo Stacho and Gabor Gevay are appreciated.

**References**

[1]   L.B. Kish, Noise-based logic: binary, multi-valued, or fuzzy, with optional superposition of logic states, Physics Letters A 373 (2009) 911–918. http://arxiv.org/abs/0808.3162

[2]   L.B. Kish, S. Khatri, S. Sethuraman, Noise-based logic hyperspace with the superposition of 2^N states in a single wire, Physics Letters A 373 (2009) 1928–1934.  http://arxiv.org/abs/0901.3947

[3]   L.B. Kish, S. Khatri, F. Peper, Instantaneous noise-based logic, Fluctuation and Noise Letters 9 (2010) 323–330.  http://arxiv.org/abs/1004.2652

[4]   F. Peper, L.B. Kish, Instantaneous, non-squeezed, noise-based logic, Fluctuation and Noise Letters 10 (2011) 231–237.  Open access.

[5]   H. Wen, L.B. Kish, Noise-based logic: Why noise? A comparative study of the necessity of randomness out of orthogonality, Fluctuation and Noise Letters 11 (2012) 1250021/1–1250021/7.  http://arxiv.org/abs/1204.2545

[6]   H. Wen, L.B. Kish, A. Klappenecker, F. Peper, New noise-based logic representations to avoid some problems with time complexity, Fluctuation and Noise Letters 11 (2012) 1250003/1–1250003/8.  http://arxiv.org/abs/1111.3859

[7] H. Wen, L.B. Kish, A. Klappenecker, Complex noise-bits and large-scale instantaneous parallel operations with low complexity, Fluctuation and Noise Letters 12 (2013) 1350002. http://vixra.org/abs/1208.0226

[8] L.L Stachó, Fast measurement of hyperspace vectors in noise-based logic, Fluctuation and Noise Letters 11 (2012) 1250001.

[9] L.B. Kish, S. Khatri, T. Horvath, Computation using noise-based logic: efficient string verification over a slow communication channel, European Journal of Physics B 79 (2011) 85–90. http://arxiv.org/abs/1005.1560

**Figures**

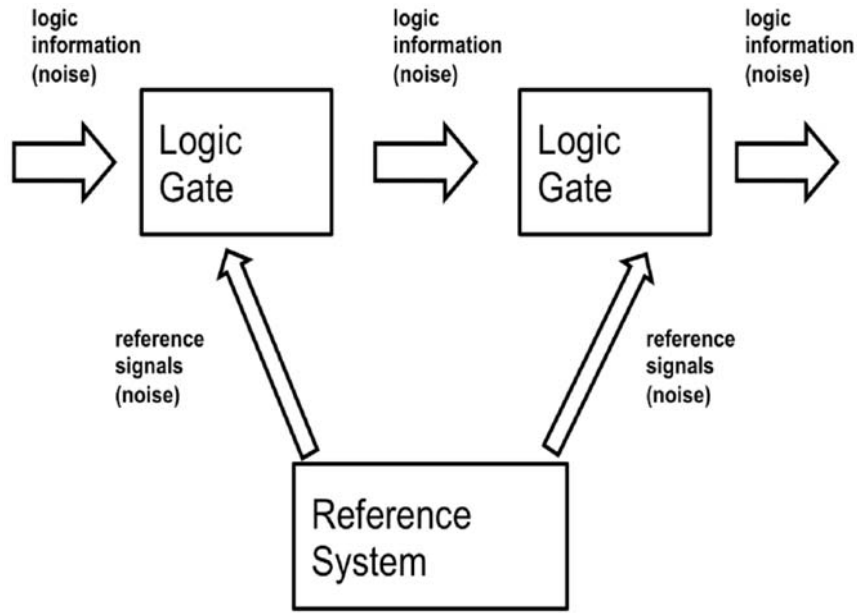

Fig. 1. Generic noise-based logic hardware scheme. In instantaneous noise-based logic, the logic gates do not use statistical correlators or time averaging [4]. Instead algebraic operations, or coincidence detection, are used to instantaneously create the output state (signal).

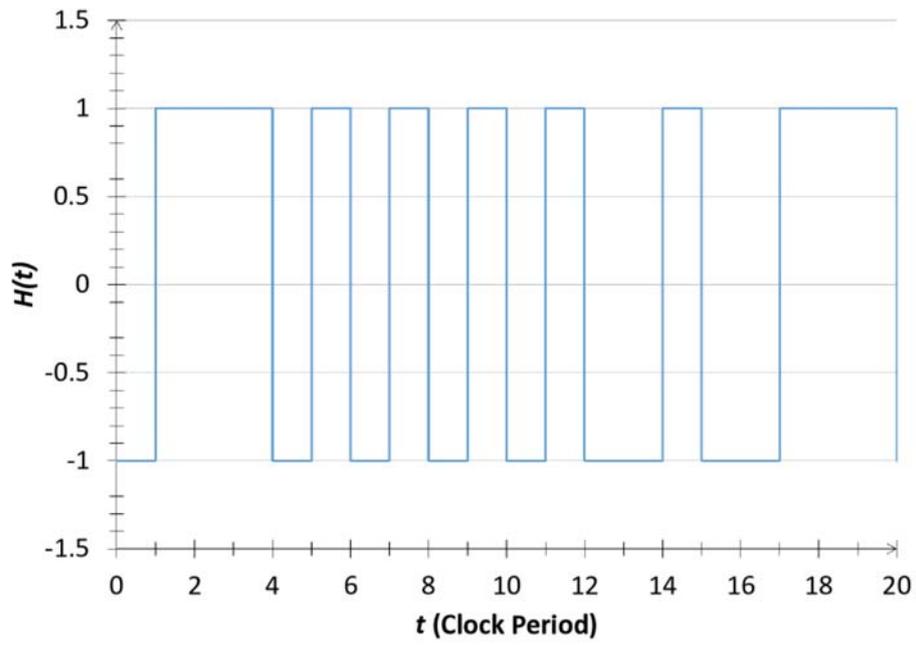

Fig. 2. Computer simulation of a random telegraph wave carrying the High-bit value of a noise-bit in the asymmetric INBL scheme.

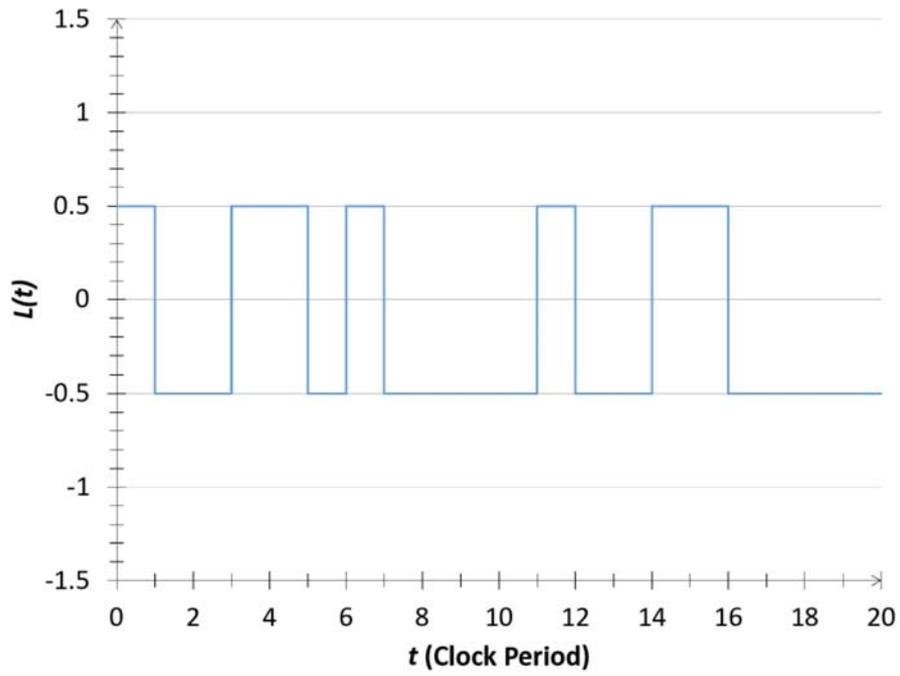

Fig. 3. Computer simulation of a random telegraph wave carrying the Low-bit value of a noise-bit in the asymmetric INBL scheme.

Fig. 4. Computer simulation of the product-string $\prod_{j=1}^{N} G_j^{(2)}(t)$ of the number 2 (binary number 00000000000000000000000000000010) in a $N = 32$ noise-bits INBL scheme.

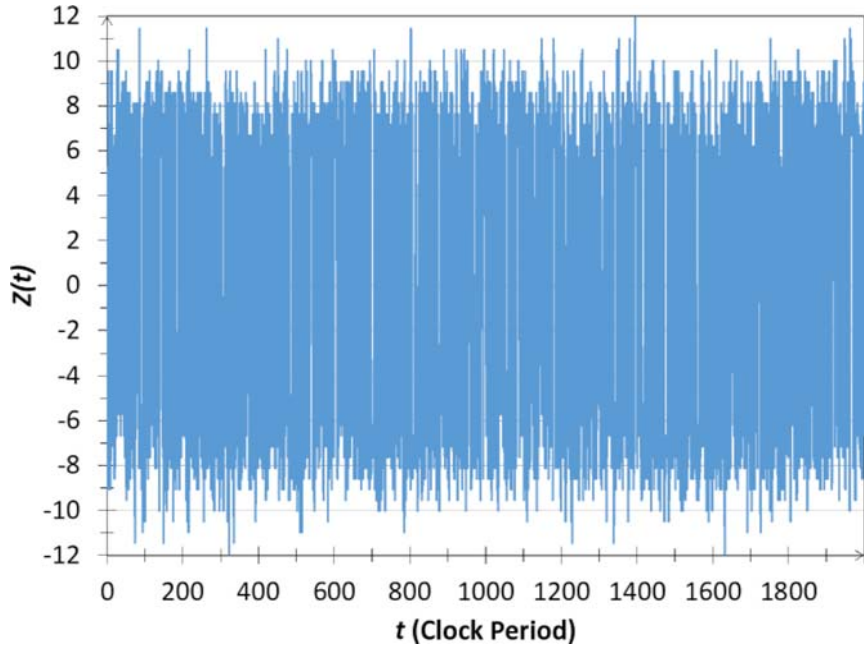

Fig. 5. Computer simulation of the *N* = 32 noise-bit universe in the asymmetric INBL scheme over 2000 clock periods. The time function $U(t)$ of universe, which represents the superposition of all integer numbers from 0 to 4294967295, is plotted in a logarithmically distorted way, defined by $Z(t) = \text{sign}[U(t)] \log[2^{32}|U(t)|]$, for better visibility of small and large variations.

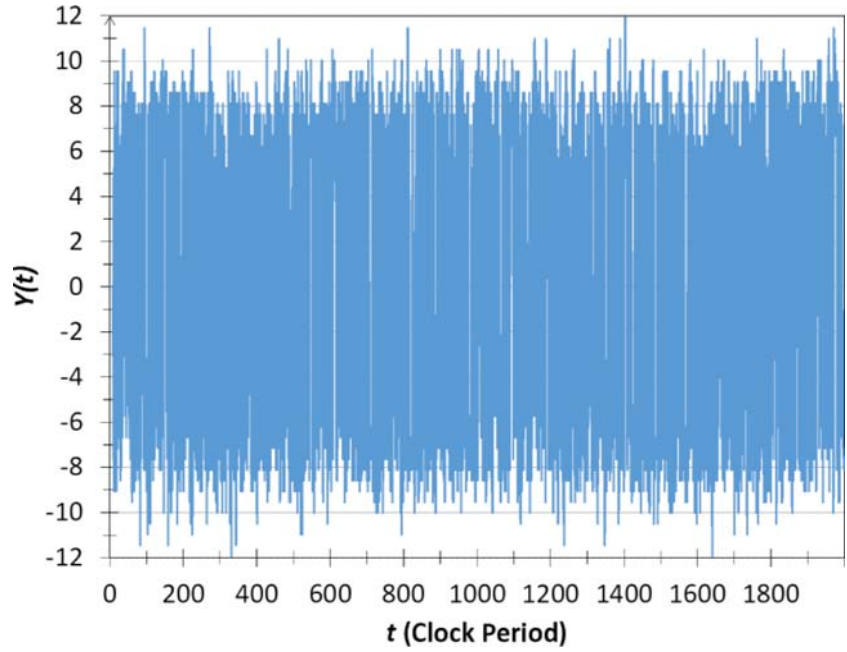

Fig. 6. Computer simulation of the *N* = 32 noise-bit universe shown in Figure 5 reduced by the product-string of the missing number, *p* = 2 (see Figure 4), where the time function $U^{(p)}(t)$ is plotted in a logarithmically distorted way, defined as $Y(t) = \text{sign}\left[U^{(p)}(t)\right] \log\left[2^{32}\left|U^{(p)}(t)\right|\right]$, for better visibility of small and large variations.

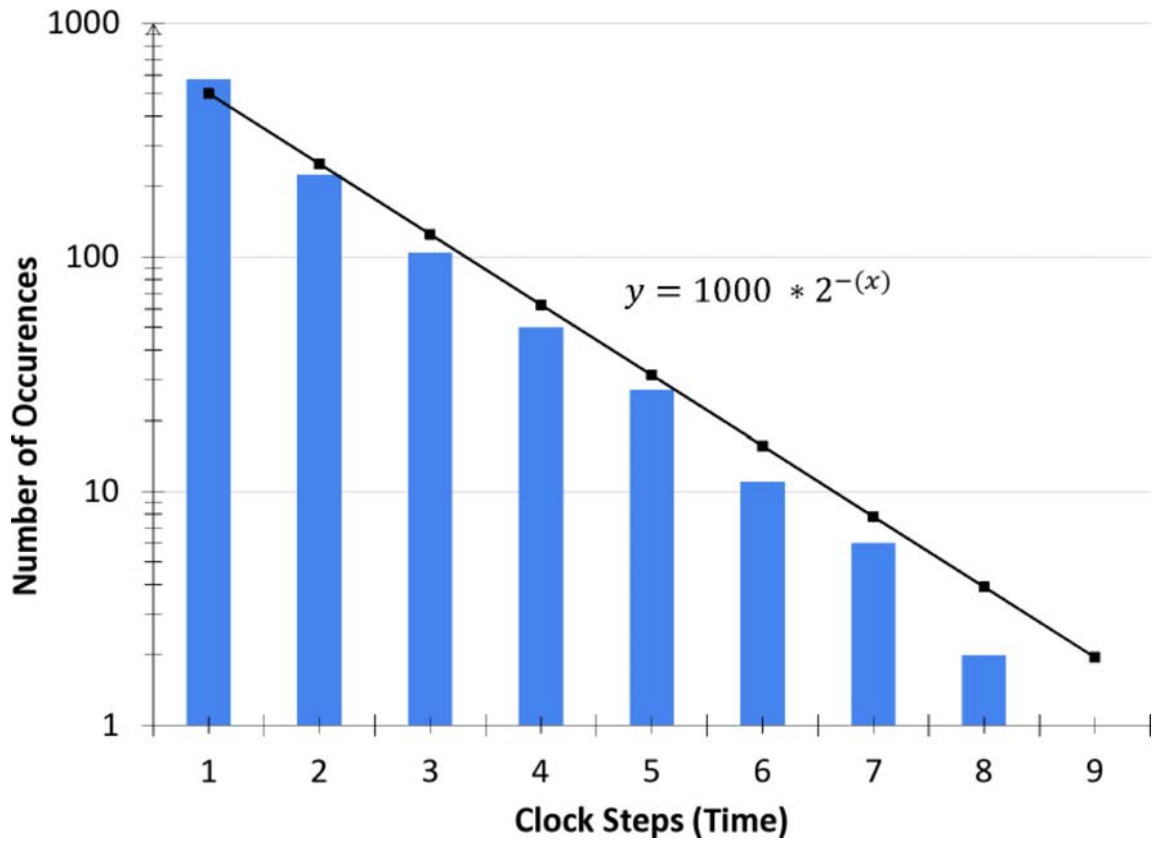

Fig. 7. The computer simulated distribution of the number of clock periods required to observe the difference of signals for an error-free response. The same $N=32$ noise bit system was used as in Figures 4-6. The numbers drawn were $p=2$ and $q=1$, that is, binary numbers 00000000000000000000000000000010 and 00000000000000000000000000000001, respectively.